\documentclass[aps,prl,floats,amsmath,twocolumn]{revtex4}
\usepackage{graphicx}
\usepackage{multirow}

\begin{document}

\title{Understanding the Unique Structural and Electronic Properties
of SrFeO$_2$}

\author{J.M.~Pruneda,$^1$ Jorge~\'I\~niguez,$^1$ Enric~Canadell,$^1$
  Hiroshi~Kageyama,$^2$ and Mikio~Takano$^3$}

\affiliation{$^1$Institut de Ci\`encia de Materials de Barcelona
(ICMAB-CSIC), Campus UAB, 08193 Bellaterra, Spain}
\affiliation{$^2$Department of Chemistry, Graduate School of Science,
Kyoto University, Sakyo, Kyoto 606-8502, Japan}
\affiliation{$^3$Institute for Chemical Research, Kyoto University,
Uji, Kyoto 611-0011, Japan}

\begin{abstract}
We report a first-principles study of SrFeO$_2$, an infinite-layer
oxide with Fe atoms in a perfect square-planar coordination down to
essentially 0~K. Our results reveal this striking behavior relies on
the double occupation of the 3$d_{z^2}$ orbitals of high-spin
Fe$^{2+}$. Such an electronic state results from the hybridization of
iron's 3$d_{z^2}$ and 4$s$ orbitals, which enables a large reduction
of the intra-atomic exchange splitting of the $z^2$ electrons. The
generality of the phenomenon is discussed.
\end{abstract}

\pacs{71.20.-b,71.70.Ch,71.15.Mb}
% 71.20.-b Electron density of states and band structure of crystalline solids
% 71.70.Ch Crystal and ligand fields
% 71.15.Mb Density functional theory, local density approximation, gradient and other corrections

\maketitle

\begin{figure}
\includegraphics[width=0.7\columnwidth]{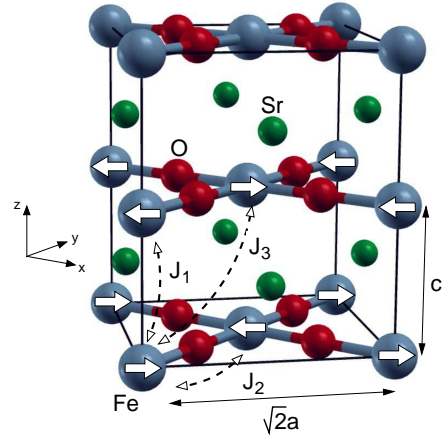}
\caption{(color online) 16-atom SrFeO$_2$ cell considered in most of
  our calculations. The ground-state magnetic structure is sketched;
  the lattice and exchange constants are defined.}
\label{fig1}
\end{figure}

Tsujimoto {\sl et al}.~\cite{tsujimoto07} have recently reported on
SrFeO$_2$, a novel material that exhibits a wealth of surprising
properties. Most interestingly, SrFeO$_2$ presents Fe$^{2+}$ ions in a
square-planar coordination, forming unprecedented infinite FeO$_2$
layers that remain undistorted down to at least 4.2~K (see
Fig.~\ref{fig1}). In addition, it undergoes a transition to an
anti-ferromagnetic (AFM) state at $T_{\rm N}$=473~K, a surprisingly
high temperature given the seemingly two-dimensional nature of the
compound. The physical origin of these features, which also appear in
similar compounds studied more recently~\cite{tassel08}, remains
unknown. Here we report a first-principles study, based on the
``LDA+U'' approach to Density Functional Theory (DFT), that confirms
the peculiar behavior of SrFeO$_2$ and reveals its electronic origin.

{\sl Basic properties}.-- We begin by showing that our DFT methods,
described in~\cite{fn-methods}, reproduce the main results reported in
Ref.~\onlinecite{tsujimoto07}. To investigate the stability of the
highly-symmetric phase of SrFeO$_2$, we first determined the
structural parameters that minimize the energy of the crystal under
the constraints that the P4/mmm space group is preserved and the
magnetic order is the AFM one found experimentally. Our results, shown
in Table~\ref{tab1}, are in good agreement with the measurements. We
then computed the elastic constants and the phonons that are
compatible with the cell of Fig.~\ref{fig1}. The results, summarized
in Table~\ref{tab1}, indicate that the P4/mmm phase of SrFeO$_2$ is
stable upon symmetry-lowering structural distortions, thus confirming
it as the ground state of the material at 0~K. It is worth noting
that, while obviously anisotropic, the computed elastic constants and
phonon frequencies present values that are acceptable for a
three-dimensional (3D) crystal (for SrFeO$_3$ we obtained
$C_{11}$=300~GPa and $C_{12}$=140~GPa), which suggests this novel
infinite-layer phase is robustly (meta-)stable.

\begin{table}
\caption{Computed properties of the P4/mmm phase of SrFeO$_2$
(measured lattice parameters in parenthesis).}
\vskip 2mm
\begin{tabular}{ll}
\hline
Structure: & $a$=$b$=4.01~\AA\ (3.99~\AA); $c$=3.42~\AA\ (3.47~\AA) \\
\multicolumn{2}{l}{Elastic constants (GPa):}\\
          & $B$=125, $C_{11}$=$C_{22}$=317, $C_{33}$=164, $C_{12}$=78, \\
          & $C_{13}$=$C_{23}$=51, $C_{44}$=$C_{55}$=95, $C_{66}$=125 \\
\multicolumn{2}{l}{Infra-red active modes (cm$^{-1}$):}\\ 
& 159 and 276 ($A_{\rm 2u}$); 183, 301, and 531 ($E_{\rm u}$) \\
\multicolumn{2}{l}
{Magnetic couplings (meV): $J_1$=1.75, J$_2$=6.58, $J_3$=$-$0.26} \\
\hline
\end{tabular}
\label{tab1}
\end{table}

We then studied the magnetic interactions in SrFeO$_2$. We considered
all the spin arrangements compatible with the cell in Fig.~\ref{fig1},
and confirmed that the experimentally observed AFM pattern (sketched
in the figure) is the magnetic ground state. Further, the Fe$^{2+}$
ions are in a high-spin configuration with a magnetic moment of about
3.6~$\mu_{\rm B}$, in agreement with experiment. We also computed the
exchange parameters of a model Hamiltonian
$H=1/2\sum_{ij}J_{ij}S_{i}S_{j}$, with $|S|$=2; the results are given
in Table~\ref{tab1}, following the notation defined in
Fig.~\ref{fig1}. We found that the in-plane interactions are
considerably stronger than those along the $c$ direction; yet, the
material is far from being two-dimensional from the magnetic point of
view. Indeed, a simple (admittedly rough) mean field estimate based on
our computed $J$'s renders a N\'eel temperature above 1000~K, and the
energy scale for the magnetic couplings between FeO$_2$ planes, which
we can estimate as $S^{2}(2J_{1}-8J_{3})$, is found to be around
260~K. Thus, our results confirm the three-dimensional character of
magnetism in SrFeO$_2$, and are compatible with the relatively high
$T_{\rm N}$ observed. We checked that values of the $U$ parameter
within the 3--6~eV range, which is typical for Fe compounds, do not
affect this conclusion.

\begin{figure}
\includegraphics[width=0.95\columnwidth]{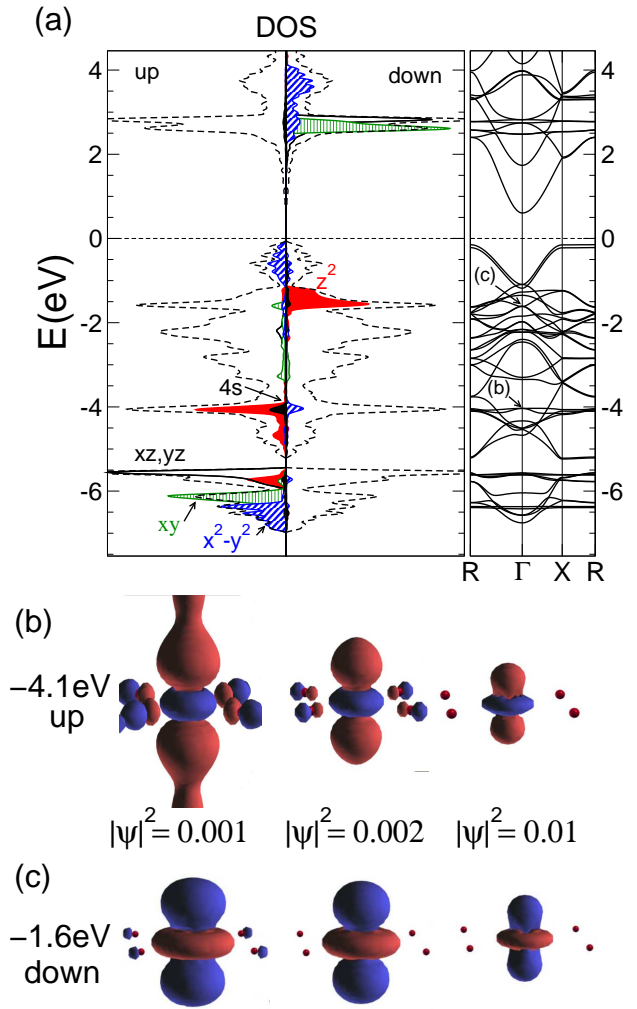}
\caption{(color online) Panel~(a): Computed electronic structure of
  SrFeO$_2$ around the Fermi energy, which is set to 0~eV. The partial
  Density of States (pDOS) corresponding to the 3$d$ and 4$s$ orbitals
  of a particular Fe atom is indicated. The dashed line depicts the
  total DOS divided by a factor of four. Panels (b) and (c):
  Representative $\Gamma$-point eigenstates with dominating $z^2$
  character, indicated with arrows in the band structure of
  panel~(a). The plotted isosurfaces correspond to the indicated
  values of the squared modulus of the wave function.}
\label{fig2}
\end{figure}

{\sl Electronic structure}.-- Based on their $^{57}$Fe M\"ossbauer
spectroscopy results showing a temperature-independent quadrupole
splitting of about 1~mm/s, and in accordance with the usual picture
from crystal-field theory, the authors of
Ref.~\onlinecite{tsujimoto07} assumed the
($xz$,$yz$)$^3$($xy$)$^1$($z^2$)$^1$($x^2-y^2$)$^1$ electronic
configuration for the 3$d$ electrons of Fe$^{2+}$. However, in such
conditions a Jahn-Teller distortion is expected to lift the degeneracy
associated to the ($xz$,$yz$) orbitals, at odds with the experimental
results. Our calculations readily clarify this question. As shown in
Fig.~\ref{fig2}, the DFT electronic structure of SrFeO$_2$ does not
agree with the one proposed in Ref.~\cite{tsujimoto07}. We found that
the sixth 3$d$ electron of high-spin Fe$^{2+}$ goes to the $z^2$
minority-spin orbital. At variance with the ($xz$,$yz$) 3$d$ orbitals
of Fe, the $z^2$ state is not degenerate with any other and, thus, its
occupancy cannot cause any symmetry-lowering distortion that reduces
the energy. Hence, this electronic configuration is compatible with
the stability of the P4/mmm phase. Further, we also checked the
calculation results are compatible with the M\"ossbauer data. We
address this point below, after having discussed the electronic
structure of SrFeO$_2$ in more detail.

{\sl Discussion}.-- Our DFT calculations provide us with a coherent
picture of SrFeO$_2$, accounting for the structural and magnetic
properties of the material. Yet, that picture relies on a result, the
double occupation of the $z^2$ orbitals of Fe$^{2+}$, which may seem
counterintuitive based on the common knowledge in molecular and
solid-state chemistry. We will thus focus on this key issue: Why are
the $z^2$ orbitals of Fe$^{2+}$ doubly occupied in SrFeO$_2$?

Let us start with a word of caution: Many apparently satisfying
answers to the above question do not resist a rigorous analysis. For
example, it may be tempting to postulate that the double occupation of
the $z^2$ orbitals is determined by the fact that, as compared with
the other 3$d$'s, they suffer from negligible Coulomb repulsion
because they find no O$^{2-}$ ions along the direction perpendicular
to the FeO$_2$ plane. This kind of argument has been invoked in
several occasions~\cite{burns66,xiang-condmat} and is supported by
extended H\"uckel calculations we have performed. However, it is
severely flawed: It implies that the $z^2$ orbitals should be the
lowest-lying ones for both the majority- and minority-spin electrons
of high-spin Fe$^{2+}$, which is in evident disagreement with the
results in Fig.~\ref{fig2}. Ultimately, the failure of this argument
suggests that an independent-electron picture is not adequate to
explain the electronic structure of SrFeO$_2$.

To address the above question, we proceeded by eliminating
possibilities in a systematic way. Our key observations are the
following. As already discussed, our calculations show SrFeO$_2$
should be regarded as a 3D crystal, and one can thus conjecture that
interactions {\sl along} the $c$ direction, particularly between
neighboring Fe atoms, might play a role in stabilizing the
doubly-occupied $z^2$ orbital. In order to investigate this
possibility, we considered the limit case in which such interactions
are totally suppressed. More precisely, we studied an idealized system
composed of an isolated FeO$_2$ layer, fixing the number of electrons
in the calculation so as to keep the Fe and O atoms in the ionization
states they have in SrFeO$_2$. We denote this virtual 2D crystal as
``Fe$^{2+}$(O$^{2-})_2$\,+\,bg'', where ``bg'' refers to the
homogeneous charge-compensating background introduced to stabilize the
material. Interestingly, the partial Density of States (pDOS) obtained
for ``Fe$^{2+}$(O$^{2-})_2$\,+\,bg'' is essentially identical to the
result for SrFeO$_2$ in Fig.~\ref{fig2}, indicating that out-of-plane
interactions play no role in determining the electronic structure of
SrFeO$_2$. Recognizing that the significance of the results with a
charge background may be questionable, we also run calculations for
SrFeO$_2$ with a significantly increased $c$ lattice constant and
ratified this conclusion.

Next, we turned our attention to the interactions within the FeO$_2$
planes. We ran calculations for slightly increased and decreased
values of the $a$=$b$ lattice constants, and found that the most
significant changes in the electronic structure pertain to the
$x^2-y^2$ electrons of iron. This is not surprising, as these are the
electrons forming $\sigma$ bonds with the oxygen atoms. In what
regards the $z^2$ electrons, we observed no effect worth noting, which
clearly indicates that the Fe--O interactions are not the key to
understand the double occupation of the $z^2$ orbitals. In particular,
these results rule out the possibility that $\pi$-donor interactions
with the {\sl oxo ligands}, which would destabilize the $xz$ and $yz$
orbitals, are at the origin of the double occupation of $z^2$. These
conclusions are ratified by a standard Fe--O overlap population
analysis. We also checked that the electronic structure of SrFeO$_2$
is essentially independent of the spin arrangement, and we can thus
neglect inter-atomic exchange effects in this discussion.

We are thus led to conclude the double occupation of the $z^2$
orbitals must be determined by some sort of {\sl intra-atomic}
mechanism associated to the iron atoms. Indeed, this is the correct
perspective to understand the electronic structure of SrFeO$_2$. Let
us note a well-known effect that appears to be critically important in
this compound: As a result of the tetragonal point symmetry at the Fe
site, the $z^2$ orbital is fully symmetric and can hybridize with the
formally-empty Fe-4$s$ state. In fact, the pDOS of Fig.~\ref{fig2}
clearly shows such a hybridization occurs, being particularly
important for the majority spin. As illustrated in panels (b) and (c)
of Fig.~\ref{fig2}, the majority-spin wave functions with a
significant 4$s$--3$d_{z^2}$ mixing are characterized by (i) a
shrinking of the $z^2$-like central ring, and (ii) a deformation, and
greater relative weight, of the lobes extending out-of-plane. In
contrast, the minority-spin electrons display a nearly pure $z^2$
character. Note also that the very different structure of the $z^2$
pDOS for the majority (roughly, triple-peaked) and minority
(single-peaked) spins is consistent with this different mixing with
the 4$s$ orbitals. Of course, this peculiar behaviour reflects a
mechanism by which the crystal minimizes its energy: Having different
spin-up and spin-down ${z^2}$-like electrons results in a reduction of
their Coulomb repulsion and, accordingly, of the corresponding
intra-atomic exchange spliting $\Delta$. Indeed, the relatively small
$\Delta$ associated to the $z^2$ electrons is obvious from
Fig.~\ref{fig2}. For a given orbital type, $\Delta$ can be estimated
as the difference between the centers of mass of the spin-up and
spin-down bands; our calculations indicate that $\Delta$ is about
2.9~eV for the $z^2$ orbitals, while the other 3$d$'s present
$\Delta$'s of more than 6~eV.

\begin{figure}
\includegraphics[width=0.7\columnwidth]{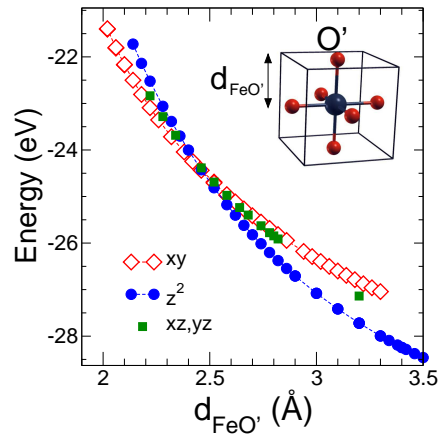}
\caption{(color online) Computed energy of various electronic
solutions of the ``Fe$^{2+}$(O$^{2-})_3$\,+\,bg'' virtual system (see
text) as a function of the distance $d_{\rm FeO'}$ between the central
Fe atom and the {\sl apical} oxygens O'.}
\label{fig3}
\end{figure}

We have thus identified the mechanism that allows SrFeO$_2$ to reduce
its energy by adopting an electronic configuration with
doubly-occupied $z^2$ orbitals. Interestingly, for this mechanism to
be active we only need to have the appropriate local symmetry at the
Fe site, which implies that the newly discovered effect might be quite
general. To check this, we performed the following computational
experiment: We considered the 3D crystal
``Fe$^{2+}$(O$^{2-})_3$\,+\,bg'' sketched in Fig.~\ref{fig3}, fixing
the number of electrons in order to retain the Fe-3$d^6$
configuration. For simplicity, we assumed a ferromagnetic order, which
we checked has no effect on the electronic configuration of Fe. We
fixed the in-plane lattice constant at the SrFeO$_2$ value, and
monitored the evolution of different electronic
solutions~\cite{fn-metastable} as the cubic crystal ($d_{\rm
FeO'}$=2~\AA) undergoes a tetragonal distortion. As shown in
Fig.~\ref{fig3}, only one electronic solution is found in the cubic
case: the minority-spin 3$d$ electron occupies a combination of
$t_{2g}$ orbitals (from which the $xy$ orbital is selected as soon as
the tetragonal distortion occurs), closing the gap of the crystal. In
contrast, for the largest values of $d_{\rm FeO'}$ considered, several
solutions occur, the ground state corresponding to doubly-occupied
$z^2$ orbitals. Interestingly, the solution with doubly-occupied $z^2$
already exists at $d_{\rm FeO'}$=2.14~\AA; yet, the ``$xy$-to-$z^2$''
crossover only occurs at $d_{\rm FeO'}$=2.50~\AA. These results show a
mere symmetry breaking will not cause the double occupation of the
$z^2$ orbitals; rather, a relatively large tetragonal distortion,
which results in both a more anisotropic environment for the Fe atom
and a reduction of the Coulomb repulsion for the $z^2$ orbitals, is
required. Finally, we also checked that changes in the in-plane
lattice constant have a minor effect in the $xy$-to-$z^2$ crossover.

It is not easy to estimate the universality of this mechanism for the
double occupation of the $z^2$ orbitals. It may seem surprising, for
example, that such an electronic solution is not found in molecules
with Fe$^{2+}$ centers. However, after reviewing the literature for
some well-known cases~\cite{molecules}, we have the impression that
the electronic structure of iron is often conditioned by the nature of
the ligands or by a reduced atomic symmetry that precedes any
electronic effect associated to the metal atom itself. This suggests
such systems may not be appropriate to look for effects like the one
discussed here. As for crystalline materials, we only know of one
candidate to present this kind of effect: the mineral gillespite
(BaFeSi$_4$O$_{10}$), which contains Fe$^{2+}$ ions in a square-planar
coordination and displaying a high-spin configuration with
doubly-occupied $z^2$ orbitals~\cite{burns66}. An appealing
possibility, though, would be to synthesize other compounds of the
SrFeO$_2$ family. Indeed, we found that crystals like RbCoO$_2$ and
YFeO$_2$ would present a similar pattern of electronic levels with
doubly-occupied $z^2$ orbitals. More precisely, our calculations
predict RbCoO$_2$ displays a Co-3$d^6$ high-spin configuration with a
doubly-occupied $z^2$; thus, this material might remain in the
high-symmetry P4/mmm phase down to 0~K, exactly as SrFeO$_2$. On the
other hand, YFeO$_2$ exhibits a Fe-3$d^7$ high-spin configuration
where, in addition to having a doubly-occupied $z^2$, the last 3$d$
electron goes to a ($xz$,$yz$) state that closes the band gap; hence,
this crystal should undergo a Jahn-Teller distortion.

Finally, let us discuss the compatibility between the value of the
quadrupole splitting (QS) obtained from $^{57}$Fe~M\"ossbauer
spectroscopy, which is about 1~mm/s, and the double-occupation of the
$z^2$ orbitals, which is expected to result in QS values of
2--3~mm/s. We should begin by noting that SrFeO$_2$ is not the only
compound exhibiting this seemingly contradictory behaviour. Indeed,
the Fe$^{2+}$ atoms in the above mentioned gillespite have been
reported to display QS values below 0.6~mm/s in the 80--650~K
temperature range~\cite{clark67}. As for SrFeO$_2$, we computed the
electric field gradient at the Fe site and the corresponding QS,
obtaining a value of about 0.8~mm/s. We also checked that varying $U$
between 0 and 6~eV renders QS results in the 0.6--0.9~mm/s range, the
greater values corresponding to the smaller $U$'s. This clearly shows
that having an electronic structure with doubly occupied $z^2$
orbitals is not incompatible with the relatively small QS values
measured. As to the mechanism that causes such a reduction of the
observed QS with respect to its expected value, two possibilities can
be mentioned: (i) that the lattice contribution to the QS is
comparable to the one coming from the 3$d$ electrons of
iron~\cite{clark67}, which would lead to a cancelation as their signs
are known to be opposite, or (ii) that the splittings associated to
iron's 3$d$ electrons are unusual in SrFeO$_2$. Interestingly, our
results seem to support the latter possibility: The spread in energy
of the majority-spin $z^2$ and $x^2-y^2$ electrons (see
Fig.~\ref{fig2}a) will imply a relatively large spacial spread of the
corresponding Wannier functions~\cite{souza01}, which might result in
unusually small quadrupole splittings. However, a detailed analysis of
this issue is not trivial and goes beyond the scope of this paper.

{\sl Summary}.-- We have used first-principles methods to show that
SrFeO$_2$, an unusual high-spin Fe$^{2+}$ layered phase recently
reported, is free from structural instabilities and exhibits strong
magnetic interactions. We find that the hybridization of iron's
3$d_{z^2}$ and 4$s$ orbitals, which results in a large reduction of
the intra-atomic exchange splitting associated to the $z^2$ electrons,
is the key feature to understand the structural stability of the
material.

We acknowledge discussions with E.~Molins. Work funded by the Catalan
(SGR2005-683) and Spanish (FIS2006-12117-C04-01, CSD2007-00041)
Governments, CSIC (PIE-200760I015), and the Marie Curie program.


\begin{thebibliography}{99}

\bibitem{tsujimoto07} Y. Tsujimoto {\sl et al}., Nature {\bf 450},
1062 (2007).

\bibitem{tassel08} C.~Tassel {\sl et al}., J. Am. Chem. Soc. (in
press).

\bibitem{fn-methods} Methods: Generalized Gradient Approximation to
DFT, as implemented in VASP within the PAW scheme~\cite{vasp};
``LDA+U'' correction~\cite{dudarev98} for Fe-3$d$ electrons, with
$U$=4~eV ($U$=0 renders qualitatively similar results); valence and
semi-core electrons considered; plane-wave cut-off: 400~eV. Most
calculations with 16-atom cell in Fig.~\ref{fig1};
4$\times$4$\times$4~$k$-point grid. We also used PWscf~\cite{pwscf}
(analysis) and ABINIT~\cite{abinit} (electric field gradient
calculations).

\bibitem{vasp} J.P.~Perdew, K.~Burke, and M.~Ernzerhof,
Phys. Rev. Lett. {\bf 77}, 3865 (1996); G. Kresse and J. Furthmuller,
Phys. Rev. B {\bf 54}, 11169 (1996); P. E. Blochl, Phys. Rev. B {\bf
50}, 17953 (1994); G. Kresse and D. Joubert, Phys. Rev. B {\bf 59},
1758 (1999).

\bibitem{dudarev98} S.L.~Dudarev {\sl et al}., Phys. Rev. B {\bf 57},
  1505 (1998).

\bibitem{pwscf} S. Baroni, A. Dal Corso, S. de Gironcoli, and
P. Giannozzi, http://www.pwscf.org/.

\bibitem{abinit} X.~Gonze {\sl et al}., Comp. Mats. Sci. {\bf 25}, 478
(2002); http://www.abinit.org/.

\bibitem{burns66} R.G.~Burns, M.G.~Clark, and A.J.~Stone,
  Inorg. Chem. {\bf 5}, 1268 (1966).

\bibitem{xiang-condmat} H.J.~Xiang, S.-H.~Wei, and M.-H.~Whangbo,
arXiv:0801.1137v1.

\bibitem{fn-metastable} Particular electronic solutions can be
  stabilized by considering specially favorable conditions (e.g., a
  large value of $d_{\rm FeO'}$ for the doubly-occupied $z^2$). The
  wave functions thus obtained can be used as the starting point of
  additional calculations, so as to complete curves as in
  Fig.~\protect\ref{fig3}.

\bibitem{molecules} O.~Kahn, {\sl Molecular Magnetism}, VCH Publ. (New
  York, 1993).

\bibitem{clark67} M.G.~Clark, G.M.~Bancroft, and A.J.~Stone,
  J. Chem. Phys. {\bf 47}, 4250 (1967).

\bibitem{souza01} I.~Souza, N.~Marzari, and D.~Vanderbilt,
Phys. Rev. B {\bf 65}, 035109 (2001).

\end{thebibliography}
\end{document}